\documentclass[aps,prl,reprint,twocolumn,superscriptaddress]{revtex4-1}
\usepackage{amsmath}
\usepackage{amsfonts}
\usepackage{amssymb}
\usepackage{amsthm}
\usepackage{graphicx}
\usepackage{CJK}
\usepackage{times}
\usepackage{mathptmx}
\usepackage[colorlinks, linkcolor=blue, urlcolor=blue, anchorcolor=blue, citecolor=blue]{hyperref}
\usepackage{mathtools}
\usepackage{lipsum}
\usepackage{array}
\usepackage[table]{xcolor}
\definecolor{mygray}{gray}{.9}

\begin{document}


\title{Determination of the Dzyaloshinskii-Moriya interactions}



\author{Jie Lu}
\email{jlu@hebtu.edu.cn}
\affiliation{College of Physics and Hebei Advanced Thin Films Laboratory, Hebei Normal University, Shijiazhuang 050024, People's Republic of China}
\author{Mei Li}
\affiliation{Physics Department, Shijiazhuang University, Shijiazhuang, Hebei 050035, People's Republic of China}
\author{X. R. Wang}
\email{phxwan@ust.hk}
\affiliation{Physics Department, The Hong Kong University of Science and Technology, Clear Water Bay, Kowloon, Hong Kong}
\affiliation{HKUST Shenzhen Research Institute, Shenzhen 518057, People's Republic of China}


\date{\today}

\begin{abstract}
Using in-plane field dependence of the precessional flow of chiral domain walls 
(DWs) to simultaneously determine bulk and interfacial Dzyaloshinskii-Moriya 
interactions (DMIs) is proposed.
It is found that effective fields of bulk and interfacial DMIs have 
respectively transverse and longitudinal components that affect differently 
the motion of chiral DWs in magnetic narrow heterostructure strips. 
The in-plane field dependence of DW velocity has a dome-shape or a canyon-shape,  
depending on whether the driving force is 
an in-plane current or an out-of-plane magnetic field.
The responses of their center shifts to the reversal of topological wall charge and current/field direction 
uniquely determine the nature and strength of DMI therein. 
Operable procedures are proposed and applied to explain existing experimental data.
\end{abstract}


\maketitle


\section{I. Introduction} 
Dzyaloshinskii-Moriya interaction (DMI), the antisymmetric exchange coupling, 
was originally proposed to explain the weak ferromagnetism in antiferromagnets\cite{Dzyaloshinsky,Moriya},
and is now known as a general interaction that widely exists in magnetic systems, especially the magnetic heterostructures.
The importance of DMI in manipulating magnetic structures and dynamics has been recognized and 
an upsurge of research was witnessed in the passing decades after a successful explanation of huge remanent magnetization enhancement due to the DMI induced by Au or Pt impurities in metallic spin glasses\cite{FertAndLevy} and distinct features of chiral DW dynamics in ultrathin magnetic films\cite{EPL_100_57002}.
The main consensuses of the community are: (i) DMI comes from the spin-orbit coupling
in magnetic systems with broken inversion symmetry either in a bulk or at an interface;
(ii) DMI is crucial for stabilizing chiral magnetic solitons, such as skyrmions\cite{Boni_Science_2009,Nagaosa_Nature_2010}
and chiral DWs\cite{Blugel_PRB_2008,ChenG_PRL_2013}; (iii) DMI plays an important role in the dynamics of both magnetic chiral solitons\cite{Linder_PRB_2017,jlu_NJP_2019} and spin waves\cite{YanPeng_PRB_2019}.
Therefore precise determination of the nature and strength of a DMI 
is not only of a fundamental issue, but also practically important.

Existing schemes for measuring the DMI strength in magnetic heterostructures all presuppose 
that only interfacial DMI (i-DMI) exists in the underlying systems. 
Generally, they belong to two groups.
Schemes in group I are based on magnetization switching (DW propagation) process\cite{Choe_PRB_2013,Adam_PRB_2016,ZhaoWeisheng_Nanoscale_2018,Diez_PRB_2019,Beach_PRB_2016,XiLi_LanzhouU_JPDAP_2018,Hayashi_PRB_2019,Koopmans_NanoLett_2016,Kuswik_JMMM_2019},
while those in group II are based on spin-wave excitation and propagation\cite{BLS_PRL_2015,BLS_Koopmans_nc_2015,BLS_Nembach_nphys_2015,YanPeng_PRApplied_2018,YanPeng_PRB_2018}.
However in real heterostructures, bulk and interfacial DMI can coexist, therefore it is urgent to 
distinguish and measure them appropriately.
In this work, we propose two parallel schemes (current-driven and field-driven) by which 
both DMIs can be simultaneously probed
via precessional flow of chiral DWs in ferromagnetic (FM) layers of narrow-strip shaped heterostructures
under in-plane magnetic fields.
The averaged wall velocities are functions of in-plane fields and the resulting curves are domes (canyons)
when walls are driven by in-plane currents (out-of-plane fields).
The responses of their center shifts to the reversal of topological wall charge 
and current/field direction uniquely determine the nature and strength of DMI therein.

\section{II. Model} 
The magnetic energy density $\mathcal{E}_{0}(\mathbf{M})$ of the FM layer in a heterostructure consists of four parts: 
the exchange part $\mathcal{E}_{\mathrm{ex}}=A(\nabla\mathbf{m})^2$ with $A$ being the exchange stiffness
and $\mathbf{m}=\mathbf{M}/M_s$ ($M_s$ represents the saturation magnetization), 
the Zeeman part $\mathcal{E}_{\mathrm{Z}}=-\mu_0\mathbf{M}\cdot\mathbf{H}_a$ with
the external applied field $\mathbf{H}_a$,
the anisotropy part $\mathcal{E}_{\mathrm{ani}}=(\mu_0 M_s^2/2)(-k_{\mathrm{E}}m_z^2+k_{\mathrm{H}}m_y^2)$ where $k_{\mathrm{E}}$ ($k_{\mathrm{H}}$) is the total (crystalline plus shape) anisotropy coefficient in easy (hard) axis,
and the DMI contribution.
For i-DMI,  $\mathcal{E}_{\mathrm{i}}=D_{\mathrm{i}}\{m_z(\mathbf{r})\nabla\cdot\mathbf{m}(\mathbf{r})-[\mathbf{m}(\mathbf{r})\cdot\nabla]m_z(\mathbf{r})\}$\cite{Bogdanov_JMMM_1994}.
While for bulk DMI (b-DMI),  $\mathcal{E}_{\mathrm{b}}=D_{\mathrm{b}}\mathbf{m}(\mathbf{r})\cdot[\nabla\times\mathbf{m}(\mathbf{r})]$\cite{Bak_JPC_1980}.
Here $D_{\mathrm{i(b)}}$ is the i(b)-DMI strength.
The corresponding DMI-induced effective fields are 
$\mathbf{H}_{\mathrm{i}}(\mathbf{r})=2D_{\mathrm{i}}[\nabla m_z-(\nabla\cdot \mathbf{m})\mathbf{e}_z]/(\mu_0 M_s)$
and 
$\mathbf{H}_{\mathrm{b}}(\mathbf{r})=-2D_{\mathrm{b}}(\nabla\times \mathbf{m})/(\mu_0 M_s)$, respectively.

\begin{figure} [htbp]
	\centering
	\includegraphics[width=0.35\textwidth]{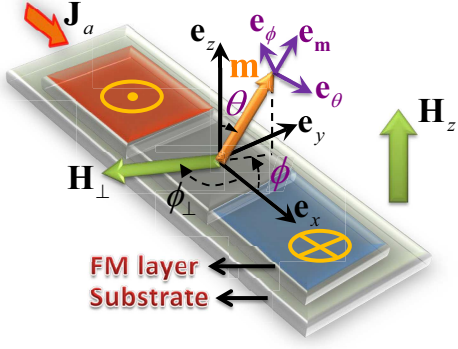}
	\caption{(Color online) Sketch of a narrow-strip shaped heterostructure in which a FM layer with 
		perpendicular magnetic anisotropy (PMA) is prepared on a nonmagnetic substrate. 
		An ``$\uparrow\downarrow$" wall is driven to move in $x-$direction 
		by external current density $\mathbf{J}_a$ or out-of-plane field $\mathbf{H}_z$. 
		Meantime an in-plane field, $\mathbf{H}_{\perp}=H_{\perp}(\cos\phi_{\perp}\mathbf{e}_x+\sin\phi_{\perp}\mathbf{e}_y)$,
		is applied. 
		($\mathbf{e}_{\mathbf{m}}\equiv\mathbf{m},\mathbf{e}_{\theta},\mathbf{e}_{\phi}$) 
		is the local spherical coordinate system associated with the magnetization unit vector $\mathbf{m}$.
	}\label{fig1}
\end{figure}

Under external field and currents, the Lagrangian $\mathcal{L}$ of this FM layer (with external normal $\mathbf{n}\equiv \mathbf{e}_z$, see Fig. 1) is
\begin{equation}\label{Lagrangian}
\frac{\mathcal{L}}{\mu_0 M_s^2}=-\frac{\cos\theta}{\gamma M_s}\frac{\partial\phi}{\partial t}-\frac{B_J\phi}{\gamma M_s}\frac{\partial\cos\theta}{\partial(\hat{\mathbf{J}}\cdot\mathbf{r})}+\frac{H_{\mathrm{FL}}}{M_s}p_{\mathbf{m}}-\frac{\mathcal{E}_{0}}{\mu_0 M_s^2},
\end{equation}
with the dissipative functional
\begin{equation}\label{Damping_functional}
\begin{split}
\frac{\mathcal{F}}{\mu_0 M_s^2}=&\frac{\alpha}{2\gamma M_s}\left\{\left[\frac{\partial}{\partial t}-\frac{\beta B_J}{\alpha}\frac{\partial}{\partial(\hat{\mathbf{J}}\cdot\mathbf{r})}\right]\mathbf{m}\right\}^2  \\ 
&\quad \quad -\frac{H_{\mathrm{ADL}}}{M_s}\left(\mathbf{m}\times\mathbf{m}_{\mathrm{p}}\right)\cdot\frac{\partial \mathbf{m}}{\partial t}. 
\end{split}
\end{equation}
describing the Gilbert damping and spin-orbit antidamping processes\cite{He_EPJB_2013,jlu_PRB_2019,Boulle_PRL_2013}.
Here $\theta(\mathbf{r},t)$ and $\phi(\mathbf{r},t)$ are the polar and 
azimuthal angles of $\mathbf{m}(\mathbf{r},t)$, respectively.
$\alpha$ is the damping constant and $\beta$ is the nonadiabatic spin-transfer torque (STT) coefficient.
$\gamma=\mu_0\gamma_e$ with $\mu_0$ and $\gamma_e$ being the vacuum permeability and electron
gyromagnetic ratio, respectively.
$B_J=\mu_{\mathrm{B}}Pj_{\mathrm{F}}/(e M_s)$, in which $\mu_{\mathrm{B}}$ is the Bohr magneton 
and $e(>0)$ is the absolute electron charge. 
$j_{\mathrm{F}}$ is the current density flowing through the FM strip 
with polarization $P$, and is usually assumed to be the same as total applied current density $j_a$ 
(with unit vector $\hat{\mathbf{J}}$) when the conductivities of FM and other layers are comparable.
$H_{\mathrm{FL}}$ and $H_{\mathrm{ADL}}$ are the strengths of field-like (FL) 
and anti-damping-like (ADL) spin-orbit torque (SOT) components, respectively.
Finally, $\mathbf{m}_{\mathrm{p}}\equiv \mathbf{n}\times\hat{\mathbf{J}}$
and is decomposed in the local ``$(\mathbf{e}_{\mathbf{m}}\equiv\mathbf{m},\mathbf{e}_{\theta},\mathbf{e}_{\phi})$"
coordinate system as $\mathbf{m}_{\mathrm{p}}=p_{\mathbf{m}}\mathbf{e}_{\mathbf{m}}+p_{\theta}\mathbf{e}_{\theta}+p_{\phi}\mathbf{e}_{\phi}$.

The magnetzation dynamics is then fully described by the Lagrangian-Rayleigh equation,
\begin{equation}\label{EL_equation}
\frac{d}{d t}\left(\frac{\delta\mathcal{L}}{\delta \dot{X}}\right)-\frac{\delta \mathcal{L}}{\delta X}+\frac{\delta\mathcal{F}}{\delta \dot{X}}=0,
\end{equation}
where an overdot means $\partial/\partial t$ and $X$ is a related coordinate.
When $X=\theta(\phi)$, the familiar Landau-Lifshitz-Gilbert equation\cite{jlu_NJP_2019} is recovered.
In principle, $\theta(\mathbf{r},t)$ and $\phi(\mathbf{r},t)$ vary from point to point,
thus generates huge number of degrees of freedom.
To obtain collective behaviors, Lagrangian-based collective coordinate models are adopted which need pre-set ansatz. 
For narrow heterostructures, the Walker ansatz\cite{Slonczewski_1972,Thiaville_EPL_2005}
\begin{equation}\label{q_phi_Delta_ansatz}
\ln\tan\frac{\vartheta}{2}=\eta\frac{x-q(t)}{\Delta},\quad \phi=\varphi(t)
\end{equation}
provides pretty good description of real wall configuration.
In this ansatz, $q$, $\Delta$ and $\varphi$ are wall center position, wall width and in-plane magnetization angle, respectively.
$\eta=+1(-1)$ corresponds to ``$\uparrow\downarrow(\downarrow\uparrow)$" wall
and is the topological wall charge.
For narrow-strip geometry as shown in Fig. 1, the $\mathbf{e}_x$ and $\mathbf{e}_y$ axes respectively 
indicate the ``longitudinal (L)" and ``transverse (T)" directions.
Accordingly, the in-plane components of effective fields from i-DMI and b-DMI are $\mathbf{H}_{\mathrm{i}}(x)=2D_{\mathrm{i}}(\nabla_x m_z)\mathbf{e}_x/(\mu_0 M_s)$ and 
$\mathbf{H}_{\mathrm{b}}(x)=2D_{\mathrm{b}}(\nabla_x m_z)\mathbf{e}_y/(\mu_0 M_s)$, respectively.
Clearly, $\mathbf{H}_{\mathrm{i(b)}}$ has longitudinal (transverse) component proportional to 
$\nabla_x m_z$, which is reversed under wall charge reversal $\eta\rightarrow -\eta$.
This leads to the totally different responses of chiral DWs under longitudinal and transverse in-plane fields.

\section{III. Current-driven $v_{\mathrm{DW}}\sim H_{\perp}$ domes}
In this section, we present the current-driven scheme.
As an example, we focus on DW dynamics under $\mathbf{J}_a=j_a\mathbf{e}_x$ in narrow heterostructure strips with pure i-DMI.
Now $H_z=0$ and an in-plane field $\mathbf{H}_{\perp}=H_{\perp}(\cos\phi_{\perp}\mathbf{e}_x+\sin\phi_{\perp}\mathbf{e}_y)$ is exerted.
By viewing $q$,$\varphi$ and $\Delta$ as three collective coordinates and integrating the resulting
dynamical equations along longitudinal direction, the following closed equation set is obtained
\begin{equation}\label{Full_dynamical_equation_jx_iDMI}
\begin{split}
\dot{q}=&-\frac{(1+\alpha\beta)B_J}{(1+\alpha^2)}-\frac{\eta\gamma\pi\Delta/2}{1+\alpha^2}\left[(H_{\mathrm{FL}}-\alpha H_{\mathrm{ADL}})\cos\varphi\right.  \\
&\left. \quad -H_{\perp}\sin(\varphi-\phi_{\perp})+\eta H_{\mathrm{i}}\sin\varphi-\frac{H_{\mathrm{K}}}{\pi}\sin 2\varphi\right], \\
\dot{\varphi}=&\frac{(\alpha-\beta)\eta B_J}{(1+\alpha^2)\Delta}+\frac{\alpha\gamma\pi/2}{1+\alpha^2}\left[(H_{\mathrm{FL}}+\frac{H_{\mathrm{ADL}}}{\alpha})\cos\varphi\right.  \\
&\left. \quad -H_{\perp}\sin(\varphi-\phi_{\perp})+\eta H_{\mathrm{i}}\sin\varphi-\frac{H_{\mathrm{K}}}{\pi}\sin 2\varphi\right], \\
\frac{\dot{\Delta}}{\Delta}=&\frac{6\gamma_0}{\alpha\pi}\left[\frac{2A}{\pi\mu_0 M_s \Delta^2}-\frac{M_s}{\pi}\left(k_{\mathrm{E}}+k_{\mathrm{H}}\sin^2\varphi\right)+H_{\mathrm{FL}}\sin\varphi\right.   \\
& \left. \quad +H_{\perp}\cos(\varphi-\phi_{\perp})\right],
\end{split}
\end{equation}
with $H_{\mathrm{K}}\equiv k_{\mathrm{H}}M_s$ and $H_{\mathrm{i}}\equiv D_{\mathrm{i}}/(\mu_0 M_s \Delta)$.
The total magnetic energy $E_0\propto \int_{-\infty}^{+\infty}\mathcal{E}_0[\mathbf{M}] dx=2A/\Delta +\mu_0 M_s^2\Delta (k_{\mathrm{E}}+k_{\mathrm{H}}\sin^2\varphi)-\pi\mu_0 M_s H_{\perp}\Delta\cos(\varphi-\phi_{\perp})+\eta\pi D_{\mathrm{i}}\cos\varphi.$
When $j_a=0$, the wall keeps static. Without $\mathbf{H}_{\perp}$,
the minimization of $E_0$ provides the static wall width $\Delta_0=\sqrt{2A/(\mu_0 k_{\mathrm{E}} M_s^2)}$
and $\cos\varphi=-\eta\mathrm{sgn}(D_{\mathrm{i}})$, 
leading to a typical N\'{e}el wall with definite chirality selected by the i-DMI.
Under finite $j_a$, the wall starts to move.
In principle, the explicit Walker limit is complicated under the coexistence of in-plane field, STT and SOT.
Nevertheless, for large enough $j_a$ the traveling-wave mode collapses and the wall falls
into the precessional-flow mode.
The time average of $\dot{q}$ gives the wall's drifting velocity.

\begin{figure*} [htbp]
	\centering
	\includegraphics[width=1.0\textwidth]{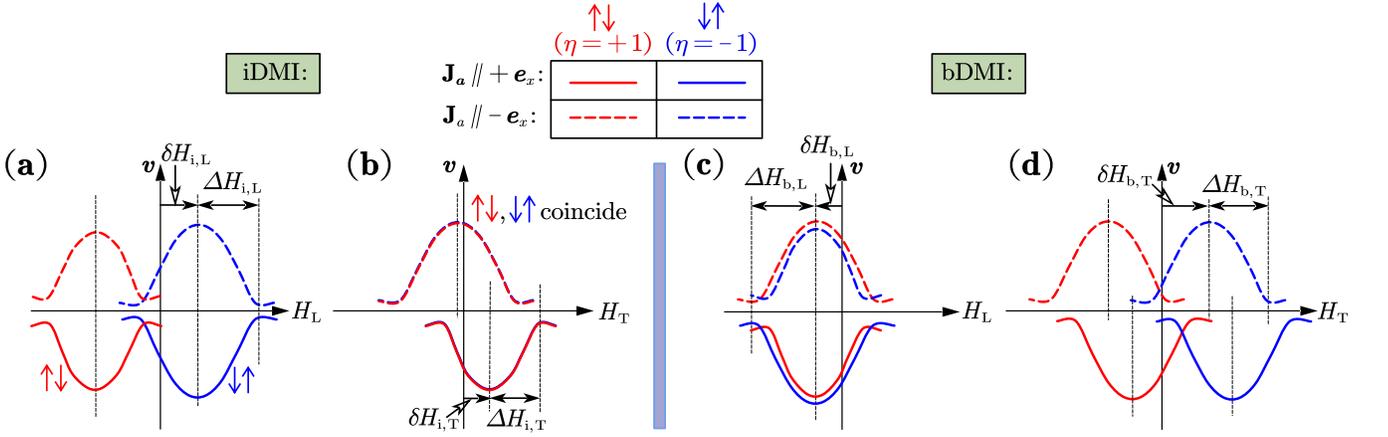}
	\caption{(Color online) Velocity dependence on $H_{\perp}$ in precessional-flow mode of current-driven DW propagation. 
		The ``weak SOT" case is taken as an example.
		\textbf{a-b}: $\uparrow\downarrow$ and $\downarrow\uparrow$ DW velocity dependence on $H_{\mathrm{L}}$ and $H_{\mathrm{T}}$ with pure i-DMI.
		$H_{\mathrm{FL(ADL)}}/j_a<0$ and $D_{\mathrm{i}}<0$ are assumed which
		corresponds to ``20 $\mathrm{\AA}$ $\mathrm{Mn}_3\mathrm{Sb}$" case in Ref.\cite{Parkin_NC_2018}.
		\textbf{c-d}: $\uparrow\downarrow$ and $\downarrow\uparrow$ DW velocity dependence on $H_{\mathrm{L}}$ and $H_{\mathrm{T}}$ with pure b-DMI.
		$H_{\mathrm{FL(ADL)}}/j_a<0$ and $D_{\mathrm{b}}<0$ are assumed.
		In all sketches, red (blue) indicates $\uparrow\downarrow$ ($\downarrow\uparrow$)
		and solid (dash) corresponds to $\mathbf{J}_a \parallel +\mathbf{e}_x(-\mathbf{e}_x)$.}\label{fig2}
\end{figure*}

\begin{table*} [htbp]
	\caption{Summary of $v\sim H_{\perp}$ domes for current-driven DW dynamics in precessional-flow mode in 
		magnetic heterostructures with PMA.
		In the combination $(\mathrm{p,q})$, $\mathrm{p}=\mathrm{i(b)}$ means i-DMI (b-DMI) dominates, 
		$\mathrm{q}=\mathrm{L(T)}$ refers to longitudinal (transverse) in-plane fields. 
		First and second row: Definition of $f_{\mathrm{p,q}}$ and $g_{\mathrm{p,q}}$.
		Third row: $v\sim H_{\perp}$ expression, in which $\tau_{\mathrm{p,q}}\equiv g_{\mathrm{p,q}}/f_{\mathrm{p,q}}$.
		Fourth row: Maximum wall velocity at dome center.
		Fifth row: Minimum wall velocity at $H_{\perp}=\delta H_{\mathrm{p,q}}\pm\Delta H_{\mathrm{p,q}}$.
		Sixth row: Dome center shifts (shaded for emphasization). 
		Last row: Half widths of domes. 
		$H_{\mathrm{i(b)}}\equiv D_{\mathrm{i(b)}}/(\mu_0 M_s \Delta)$ and 
		$H_{\mathrm{SOT}}\equiv H_{\mathrm{FL}}+H_{\mathrm{ADL}}/\alpha$.}
	\renewcommand\arraystretch{2.0}
	\begin{tabular} {p{38 pt}<{\centering} | p{105 pt}<{\centering} | p{105 pt}<{\centering} | p{130 pt}<{\centering} | p{110 pt}<{\centering}}
		\hline
		\hline
		$(\mathrm{p,q})$:  & $(\mathrm{i,L})$ & $(\mathrm{i,T})$ & $(\mathrm{b,L})$ & $(\mathrm{b,T})$   \\
		\hline
		$f_{\mathrm{p,q}}$: & $\frac{(\alpha-\beta) B_J}{\Delta} + \frac{\alpha\pi\gamma}{2}\eta H_{\mathrm{SOT}}$ & $\frac{(\alpha-\beta) B_J}{\Delta} +\frac{\alpha\pi\gamma}{2}  H_{\mathrm{i}}$ & $\frac{(\alpha-\beta) B_J}{\Delta} + \frac{\alpha\pi\gamma}{2}\left(\eta H_{\mathrm{SOT}}- H_{\mathrm{b}}\right)$ & $\frac{(\alpha-\beta) B_J}{\Delta}$   \\
		\hline
		$\frac{2}{\alpha\pi\gamma}\cdot g_{\mathrm{p,q}}$:  & $H_{\mathrm{L}}+\frac{2}{\pi}H_{\mathrm{K}}-\eta H_{\mathrm{i}}$ & $H_{\mathrm{T}}+H_{\mathrm{SOT}}-\frac{2}{\pi}H_{\mathrm{K}}$ & $H_{\mathrm{L}}+\frac{2}{\pi}H_{\mathrm{K}}$ & $H_{\mathrm{T}}+H_{\mathrm{SOT}}-\frac{2}{\pi}H_{\mathrm{K}}-\eta H_{\mathrm{b}}$  \\
		\hline
		$v_{\mathrm{p,q}}$: &  Eq. (\ref{v_iDMI_jx_Hx}) & Eq. (\ref{v_iDMI_jx_Hy}) & Eq. (\ref{v_iDMI_jx_Hx}) with $\tau_{\mathrm{b,L}}$ &  Eq. (\ref{v_iDMI_jx_Hy}) with $\tau_{\mathrm{b,T}}$ \\
		\hline
		$v_{\mathrm{p,q}}^{\mathrm{M}}$: & $-\frac{1+\alpha\beta}{1+\alpha^2} B_J - \frac{\pi\Delta\gamma}{2(1+\alpha^2)} \eta H_{\mathrm{SOT}}$  & $-\frac{1+\alpha\beta}{1+\alpha^2} B_J - \frac{\pi\Delta\gamma}{2(1+\alpha^2)}H_{\mathrm{i}}$  & $-\frac{1+\alpha\beta}{1+\alpha^2} B_J - \frac{\pi\Delta\gamma}{2(1+\alpha^2)}\left(\eta H_{\mathrm{SOT}}- H_{\mathrm{b}}\right)$ & $-\frac{1+\alpha\beta}{1+\alpha^2} B_J$  \\
		\hline
		$v_{\mathrm{p,q}}^{\mathrm{m}}$: & $-\frac{\beta}{\alpha}B_J+\eta\frac{\pi\Delta\gamma}{2\alpha}H_{\mathrm{ADL}}$ & $-\frac{\beta}{\alpha}B_J-\mathrm{sgn}(\tau_{\mathrm{i,T}})\frac{\pi\Delta\gamma}{2\alpha}H_{\mathrm{ADL}}$ & $-\frac{\beta}{\alpha}B_J+\eta\frac{\pi\Delta\gamma}{2\alpha}H_{\mathrm{ADL}}$ & $-\frac{\beta}{\alpha}B_J-\mathrm{sgn}(\tau_{\mathrm{b,T}})\frac{\pi\Delta\gamma}{2\alpha}H_{\mathrm{ADL}}$   \\
		\hline
		\rowcolor{mygray}[1.82pt][1.82pt]  
		$\delta H_{\mathrm{p,q}}$: & $\eta H_{\mathrm{i}}-\frac{2}{\pi}H_{\mathrm{K}}$ & $\frac{2}{\pi}H_{\mathrm{K}}-H_{\mathrm{SOT}}$ & $-\frac{2}{\pi}H_{\mathrm{K}}$ & $\eta H_{\mathrm{b}}+\frac{2}{\pi}H_{\mathrm{K}}-H_{\mathrm{SOT}}$ \\
		\hline
		$\Delta H_{\mathrm{p,q}}$: & \multicolumn{4}{c}{$\frac{2}{\alpha\pi\gamma}|f_{\mathrm{p,q}}|$}  \\
		\hline
		\hline
	\end{tabular}
\end{table*}

First we consider longitudinal in-plane fields ($\mathbf{H}_{\perp}=H_{\mathrm{L}}\mathbf{e}_x$).
For large enough currents, after performing linearization of $\sin\varphi$ and $\sin 2\varphi$ for $|\varphi| < 1$,
the second equation in (\ref{Full_dynamical_equation_jx_iDMI}) turns to
$(1+\alpha^2)\dot{\varphi}=\eta f_{\mathrm{i,L}}-g_{\mathrm{i,L}}\cdot\varphi$,
with $f_{\mathrm{i,L}}\equiv (\alpha-\beta) B_J/\Delta +\eta\alpha\pi\gamma(H_{\mathrm{FL}}+H_{\mathrm{ADL}}/\alpha)/2$ 
and $g_{\mathrm{i,L}}\equiv \alpha\pi\gamma(H_{\mathrm{L}}+2 H_{\mathrm{K}}/\pi-\eta H_{\mathrm{i}})/2$.
When $g_{\mathrm{i,L}}=0$ (i.e. $H_{\mathrm{L}}=\delta H_{\mathrm{i,L}}\equiv \eta H_{\mathrm{i}}-2 H_{\mathrm{K}}/\pi$), 
the wall rotates evenly ($\dot{\varphi}=const$) thus leading to constant velocity
$v_{\mathrm{i,L}}^{\mathrm{M}}=-[(1+\alpha\beta)B_J+\eta\pi\Delta\gamma(H_{\mathrm{FL}}+H_{\mathrm{ADL}}/\alpha)/2]/(1+\alpha^2)$.
When $g_{\mathrm{i,L}}\ne 0$,
$\varphi(t)=(\eta f_{\mathrm{i,L}}/g_{\mathrm{i,L}})\cdot\{1-\exp[-g_{\mathrm{i,L}} t/(1+\alpha^2)]\}$
and $\dot{q}=-\beta B_J/\alpha+\eta\pi\Delta\gamma H_{\mathrm{ADL}}/(2\alpha)-[\Delta\alpha^{-1}f_{\mathrm{i,L}} /(1+\alpha^2)]\cdot\exp[-g_{\mathrm{i,L}} t/(1+\alpha^2)]$.
The time needed for $\varphi$ changing from 0 to $\eta\mathrm{sgn}(f_{\mathrm{i,L}})$ is
$\delta t=-(1+\alpha^2)|g_{\mathrm{i,L}}|^{-1}\ln(1-|g_{\mathrm{i,L}}/f_{\mathrm{i,L}}|)$.
By defining $\tau_{\mathrm{i,L}}\equiv g_{\mathrm{i,L}}/f_{\mathrm{i,L}}$, the average wall velocity then reads
\begin{equation}\label{v_iDMI_jx_Hx}
v_{\mathrm{i,L}}=\frac{\eta\pi\Delta\gamma H_{\mathrm{ADL}}}{2\alpha}-\frac{\beta}{\alpha}B_J+\frac{\Delta\cdot f_{\mathrm{i,L}}}{\alpha(1+\alpha^2)}\cdot\frac{|\tau_{\mathrm{i,L}}|}{\ln\left(1-|\tau_{\mathrm{i,L}}|\right)}
\end{equation}
with the constraint $|\tau_{\mathrm{i,L}}|<1$.
Clearly it achieves its extremum $v^{\mathrm{m}}_{\mathrm{L}}=\eta\pi\Delta\gamma H_{\mathrm{ADL}}/(2\alpha)-\beta B_J/\alpha$ 
when $\ln(1-|\tau_{\mathrm{i,L}}|)\rightarrow 0$, that is $H_{\mathrm{L}}=\delta H_{\mathrm{i,L}}\pm \Delta H_{\mathrm{i,L}}$ with $\Delta H_{\mathrm{i,L}}=2|f_{\mathrm{i,L}}|/(\alpha\pi\gamma)$.
In real magnetic heterostructures, the effective damping in FM strips is enhanced from
$0.001\sim 0.01$ to $0.2\sim 0.9$\cite{Marrows_PRB_2018,Pizzini_EPL_2016}. Meantime $\beta$ remains the order of 0.01.
Consequently, the $v_{\mathrm{i,L}}\sim H_{\mathrm{L}}$ curve is a symmetric dome 
with respect to $H_{\mathrm{L}}=\delta H_{\mathrm{i,L}}$. 
In ``weak SOT" limit, the two ``$\eta=\pm 1$" domes for $j_a>0$ (thus $B_J>0$) locates in $v<0$ half plane.
This corresponds to ``20 $\mathrm{\AA}$ $\mathrm{Mn}_3\mathrm{Sb}$" case in Ref.\cite{Parkin_NC_2018}.
In addition, all four $v_{\mathrm{i,L}}\sim H_{\mathrm{L}}$ domes ($\eta=\pm 1$, $\mathbf{J}_a\parallel \pm \mathbf{e}_x$) 
are fully nondegenerate due to the presence of i-DMI effective field in longitudinal direction [see Fig. 2(a)].
As SOT increases, for appropriate combination of $(H_{\mathrm{FL}},H_{\mathrm{ADL}})$, the 
$v_{\mathrm{i,L}}\sim H_{\mathrm{L}}$ domes for $j_a>0$ can be reversed up to $v>0$ half plane.
The latest examples are ``20 $\mathrm{\AA}$ $\mathrm{Mn}_3\mathrm{Ge}$" and ``10 $\mathrm{\AA}$ $\mathrm{Mn}_3\mathrm{Sn}$"
cases in Ref.\cite{Parkin_NC_2018}.

For transverse in-plane fileds ($\mathbf{H}_{\perp}=H_{\mathrm{T}} \mathbf{e}_y$), 
the resulting $v_{\mathrm{i,T}}\sim H_{\mathrm{T}}$ curve is also dome-shaped with its center 
locating at $\delta H_{\mathrm{i,T}}=2H_{\mathrm{K}}/\pi-H_{\mathrm{FL}}-H_{\mathrm{ADL}}/\alpha$
hosting a maximum wall velocity $-[(1+\alpha\beta)B_J+\pi\Delta\gamma H_{\mathrm{i}}/2]/(1+\alpha^2)$. 
Following similar procedure, the average wall velocity is
\begin{equation}\label{v_iDMI_jx_Hy}
\begin{split}
v_{\mathrm{i,T}}=& -\frac{\beta}{\alpha}B_J-\frac{\pi\Delta\gamma H_{\mathrm{ADL}}}{2\alpha}\left[\frac{1}{\tau_{\mathrm{i,T}}}+\frac{\mathrm{sgn}(\tau_{\mathrm{i,T}})}{\ln(1-|\tau_{\mathrm{i,T}}|)}\right]   \\
& \qquad \quad  +\frac{\Delta\cdot f_{\mathrm{i,T}}}{\alpha(1+\alpha^2)}\cdot\frac{|\tau_{\mathrm{i,T}}|}{\ln(1-|\tau_{\mathrm{i,T}}|)},\quad |\tau_{\mathrm{i,T}}|<1
\end{split}
\end{equation}
with $\tau_{\mathrm{i,T}}\equiv g_{\mathrm{i,T}}/f_{\mathrm{i,T}}$, $f_{\mathrm{i,T}}\equiv (\alpha-\beta) B_J/\Delta +\alpha\pi\gamma H_{\mathrm{i}}/2$ 
and $g_{\mathrm{i,T}}\equiv \alpha\pi\gamma(H_{\mathrm{T}}+H_{\mathrm{FL}}+H_{\mathrm{ADL}}/\alpha-2 H_{\mathrm{K}}/\pi)/2$.
The minimum velocity, $-\beta B_J/\alpha-\pi\Delta\gamma H_{\mathrm{ADL}}\mathrm{sgn}(\tau_{\mathrm{i,T}})/(2\alpha)$, is achieved at $H_{\mathrm{T}}=\delta H_{\mathrm{i,T}}\pm\Delta H_{\mathrm{i,T}}$ with $\Delta H_{\mathrm{i,T}}=2|f_{\mathrm{i,T}}|/(\alpha\pi\gamma)$.
As shown in Fig. 2(b), now the center shifts of two $v_{\mathrm{i,T}}\sim H_{\mathrm{T}}$ domes ($\eta=\pm 1$) 
under the same current coincide 
due to the absence of $\mathbf{H}_{\mathrm{i}}$ component in transverse direction.
Physically, the center shifts of all these domes
come from the total internal effective fields in the corresponding direction.
When completely balanced by external in-plane fields, the wall rotates almost evenly thus reaches its extremum velocity.

Parallel analytics can be done when b-DMI dominates.
The resulting dynamical equation set is the same as Eq. (\ref{Full_dynamical_equation_jx_iDMI}),
except for the substitution ``$D_{\mathrm{i}}\sin\varphi\rightarrow -D_{\mathrm{b}}\cos\varphi$".
Accordingly, the $v_{\mathrm{b,L(T)}}\sim H_{\mathrm{L(T)}}$ curves
are also dome-shaped, however response differently to the reversal of $\eta$ 
and $\mathbf{J}_a$ due to the different definitions of $f_{\mathrm{b,L(T)}}$ and
$g_{\mathrm{b,L(T)}}$[see Figs. 2(c)-2(d) and the corresponding columns in Table I].
Similarly, the center shifts of these domes stem from the total internal effective fields in the corresponding axes.

Given the results above, we propose the following procedure to simultaneously probe both DMIs in 
a narrow-strip shaped heterostructure:

\noindent
(C1) Prepare quasi 1D DWs in FM layer with different topological charge ($\eta=\pm 1$).

\noindent
(C2) Apply a strong enough (exceeding Walker limit) in-plane current $\mathbf{J}_a$ along $\mathbf{e}_x$. 
For each $\eta$, the dependence of wall drifting velocity on $H_{\perp}=H_{\mathrm{L}}$ is measured. 
Reverse the current direction with unchanged strength and repeat the measurements. 
Then four $v\sim H_{\mathrm{L}}$ domes ($\eta=\pm 1, \mathbf{J}_a\parallel \pm\mathbf{e}_x$) are obtained. 

\noindent
(C3) Repeat the measurements in steps (C2) for $H_{\perp}=H_{\mathrm{T}}$ to obtain another four $v\sim H_{\mathrm{T}}$ domes.

\noindent
(C4) For fixed $\mathbf{J}_a$, if the center shifts $\delta H_{\mathrm{L}}$ of $v\sim H_{\mathrm{L}}$ domes split when $\eta=+1\rightarrow -1$, 
then i-DMI exists with strength $D_{\mathrm{i}}=\mu_0 M_s \Delta_0\cdot\left[(\delta H_{\mathrm{L}})_{\eta=+1}-(\delta H_{\mathrm{L}})_{\eta=-1}\right]/2$. 
The justification of using static DW width $\Delta_0$ instead of $\Delta$ is similar to
Appendix B of Ref.\cite{Thiaville_PRB_2016}.

\noindent
(C5) For fixed $\mathbf{J}_a$, if the center shifts $\delta H_{\mathrm{T}}$ of $v\sim H_{\mathrm{T}}$ domes split when $\eta=+1\rightarrow -1$, 
then b-DMI exists with strength $D_{\mathrm{b}}=\mu_0 M_s \Delta_0\cdot\left[(\delta H_{\mathrm{T}})_{\eta=+1}-(\delta H_{\mathrm{T}})_{\eta=-1}\right]/2$.

The above procedure and related physics can be perfectly applied to the newly released experimental data 
in unit-cell-thick perpendicularly magnetized Heusler films\cite{Parkin_NC_2018}.
First, for all three materials ($\mathrm{Mn}_3\mathrm{Ge}$, $\mathrm{Mn}_3\mathrm{Sn}$ and $\mathrm{Mn}_3\mathrm{Sb}$) therein, 
center shifts of $v\sim H_{\mathrm{L}}$ domes split when $\eta=+1\rightarrow -1$,
declaring the existence of finite i-DMI. 
Meantimes, all center shifts of $v\sim H_{\mathrm{T}}$ domes coincide for $\eta=\pm 1$ under the fixed current, thus
excludes the possibility of finite b-DMI.
In addition, from the center shifts of $v\sim H_{\mathrm{L}}$ domes and wall widths obtained already,
i-DMI strengths for 20 $\mathrm{\AA}$ $\mathrm{Mn}_3\mathrm{Ge}$, 10 $\mathrm{\AA}$ $\mathrm{Mn}_3\mathrm{Sn}$
and 20 $\mathrm{\AA}$ $\mathrm{Mn}_3\mathrm{Sb}$
films are estimated as $-0.5$ $\mathrm{mJ\; m^{-2}}$, $12.5$ $\mathrm{mJ\; m^{-2}}$ and $-2.88$ $\mathrm{mJ\; m^{-2}}$, respectively.
Second, for $\mathbf{J}_a\parallel +\mathbf{e}_x$ and $\eta=+1$, the $v\sim H_{\mathrm{L}}$ domes for 20 $\mathrm{\AA}$ $\mathrm{Mn}_3\mathrm{Sb}$
lie in $v<0$ half plane, while those for 20 $\mathrm{\AA}$ $\mathrm{Mn}_3\mathrm{Ge}$ and 10 $\mathrm{\AA}$ $\mathrm{Mn}_3\mathrm{Sn}$ lie in $v>0$ half plane. 
This indicates that SOTs in the latter two materials are stronger than that in the former, so that
the $v\sim H_{\mathrm{L}}$ domes are reversed up.
Third, the original exclusion of i-DMI in that work by the unchanged 
center shifts of $v\sim H_{\mathrm{L}}$ domes for 10 $\mathrm{\AA}$ $\mathrm{Mn}_3\mathrm{Sn}$ 
with additional CoGa overlayer capped is questionable.
Indeed, i-DMI describes the exchange interaction between magnetization in Heusler films
intermediated by heavy-metal atoms in Ta substrates.
Thus it should not be affected too much when the CoGa overlayer is added.
On the other hand, we cautiously assume that the wall width does not vary much after CoGa is capped.
Therefore the nearly unchanged $\delta H_{\mathrm{L}}$ is understandable.
At last, the shrink of wall velocity can be attributed to further shunting of total current by additional layers.

\section{IV. Field-driven $v_{\mathrm{DW}}\sim H_{\perp}$ canyons}
Except for the current-driven scheme in the above section, the field-driven counterpart can also be proposed.
Now the motion of chiral DWs is induced by pure out-of-plane field $\mathbf{H}_z$,
thus $B_J$, $H_{\mathrm{FL}}$ and $H_{\mathrm{ADL}}$ are all absent. 
For i-DMI, the closed equation set turns to
\begin{equation}\label{Full_dynamical_equation_Hz_iDMI}
\begin{split}
\dot{q}=&\frac{\eta\gamma\pi\Delta/2}{1+\alpha^2}\left[\frac{2\alpha}{\pi}H_z+H_{\perp}\sin(\varphi-\phi_{\perp})\right.  \\
&\left. \qquad \qquad \quad -\eta H_{\mathrm{i}}\sin\varphi+\frac{H_{\mathrm{K}}}{\pi}\sin 2\varphi\right], \\
\dot{\varphi}=&\frac{\alpha\gamma\pi/2}{1+\alpha^2}\left[\frac{2}{\alpha\pi}H_z-H_{\perp}\sin(\varphi-\phi_{\perp})\right.  \\
&\left. \qquad \qquad \quad +\eta H_{\mathrm{i}}\sin\varphi-\frac{H_{\mathrm{K}}}{\pi}\sin 2\varphi\right], \\
\frac{\dot{\Delta}}{\Delta}=&\frac{6\gamma_0}{\alpha\pi}\left[\frac{2A}{\pi\mu_0 M_s \Delta^2}-\frac{M_s}{\pi}\left(k_{\mathrm{E}}+k_{\mathrm{H}}\sin^2\varphi\right)\right.  \\
&\left. \qquad \qquad \quad +H_{\perp}\cos(\varphi-\phi_{\perp})\right].
\end{split}
\end{equation}
For longitudinal in-plane fields ($\mathbf{H}_{\perp}=H_{\mathrm{L}}\mathbf{e}_x$),
the second equation in (\ref{Full_dynamical_equation_Hz_iDMI}) turns to
$(1+\alpha^2)\dot{\varphi}=\tilde{f}_{\mathrm{i,L}}-\tilde{g}_{\mathrm{i,L}}\cdot\varphi$, with
$\tilde{f}_{\mathrm{i,L}}\equiv \gamma H_z$
and 
$\tilde{g}_{\mathrm{i,L}}\equiv\alpha\pi\gamma(H_{\mathrm{L}}-\eta H_{\mathrm{i}}+2 H_{\mathrm{K}}/\pi)/2$.
The average wall velocity then reads 
\begin{equation}\label{v_iDMI_Hz_Hy}
v_{\mathrm{i,L}}=\eta\frac{\Delta}{\alpha}\gamma H_z+\frac{\eta\Delta\cdot \tilde{f}_{\mathrm{i,L}}}{\alpha(1+\alpha^2)}\cdot\frac{|\tilde{g}_{\mathrm{i,L}}/\tilde{f}_{\mathrm{i,L}}|}{\ln\left(1-|\tilde{g}_{\mathrm{i,L}}/\tilde{f}_{\mathrm{i,L}}|\right)}.
\end{equation}
Generally $\alpha<1$, thus the $v_{\mathrm{i,L}}\sim H_{\mathrm{L}}$ curve is canyon-shaped, as shown in Fig. 3(a).
Its minimum, $\eta\alpha\Delta\gamma H_z/(1+\alpha^2)$, locates at $H_{\mathrm{L}}=\delta H_{\mathrm{i,L}}=\eta H_{\mathrm{i}}-2 H_{\mathrm{K}}/\pi$.
While at $H_{\mathrm{L}}=\delta H_{\mathrm{i,L}}\pm\Delta H_{\mathrm{i,L}}$ with $\Delta H_{\mathrm{i,L}}=2|\tilde{f}_{\mathrm{i,L}}|/(\alpha\pi\gamma)$ the velocity reaches its maximum $\eta\Delta\gamma H_z/\alpha$.
Parallel deductions are performed for other three cases and the resulting $v_{\mathrm{i,T}}\sim H_{\mathrm{T}}$ and  
$v_{\mathrm{b,L(T)}}\sim H_{\mathrm{L(T)}}$ curves take the similar form as in
Eq. (\ref{v_iDMI_Hz_Hy}) thus are also canyon-shaped [see Figs. 3(b)-3(d)].
The corresponding center shifts and half width are listed in Table II.

\begin{figure} [htbp]
	\centering
	\includegraphics[width=0.46\textwidth]{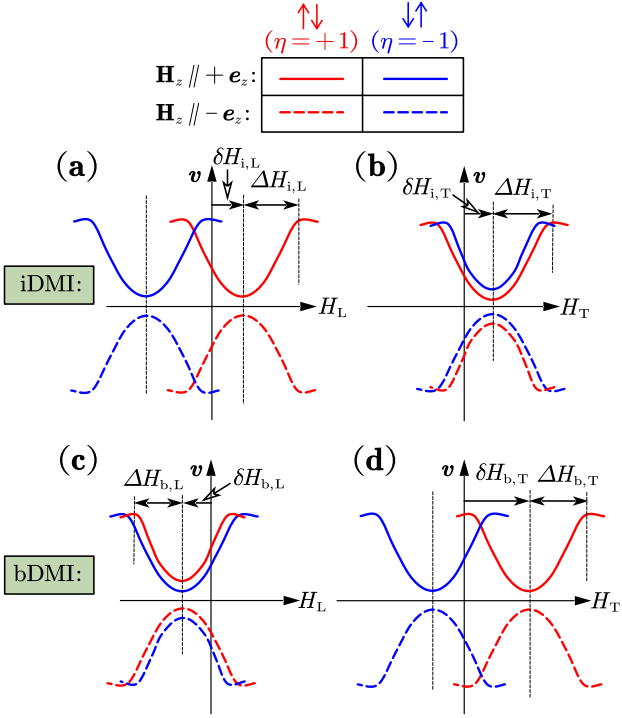}
	\caption{(Color online) Velocity dependence on $H_{\perp}$ in precessional-flow mode of field-driven DW propagation. 
		\textbf{a-b}: $\uparrow\downarrow$ and $\downarrow\uparrow$ DW velocity dependence on $H_{\mathrm{L}}$ and $H_{\mathrm{T}}$ under pure i-DMI.
		\textbf{c-d}: $\uparrow\downarrow$ and $\downarrow\uparrow$ DW velocity dependence on $H_{\mathrm{L}}$ and $H_{\mathrm{T}}$ under pure b-DMI.
		Red (blue) indicates $\uparrow\downarrow$ ($\downarrow\uparrow$)
		and solid (dash) corresponds to $\mathbf{H}_z\parallel +\mathbf{e}_z(-\mathbf{e}_z)$.}\label{fig3}
\end{figure}

Based on these results, similar procedure of probing DMIs in narrow heterostructure strips using out-of-plane fields can be proposed:

\noindent
(F1) Prepare quasi 1D DWs in FM layer with different topological charge ($\eta=\pm 1$).

\noindent
(F2) Apply a strong enough (exceeding Walker limit) out-of-plane field $\mathbf{H}_z$. 
For each $\eta$, the dependence of wall drifting velocity on $H_{\mathrm{L}}$ is measured. 
Reverse the direction of $\mathbf{H}_z$ with unchanged strength and repeat the measurements. 
Then four $v\sim H_{\mathrm{L}}$ canyons ($\eta=\pm 1, \mathbf{H}_z\parallel \pm\mathbf{e}_z$) are obtained. 

\noindent
(F3) Repeat the measurements in steps (F2) for $H_{\mathrm{T}}$ to obtain another four $v\sim H_{\mathrm{T}}$ canyons.

\noindent
(F4) For fixed $\mathbf{H}_z$, if the center shifts $\delta H_{\mathrm{L}}$ of $v\sim H_{\mathrm{L}}$ canyons split when $\eta=+1\rightarrow -1$, 
then i-DMI exists with strength $D_{\mathrm{i}}=\mu_0 M_s \Delta_0\cdot\left[(\delta H_{\mathrm{L}})_{\eta=+1}-(\delta H_{\mathrm{L}})_{\eta=-1}\right]/2$.

\noindent
(F5) For fixed $\mathbf{H}_z$, if the center shifts $\delta H_{\mathrm{T}}$ of $v\sim H_{\mathrm{T}}$ canyons split when $\eta=+1\rightarrow -1$, 
then b-DMI exists with strength $D_{\mathrm{b}}=\mu_0 M_s \Delta_0\cdot\left[(\delta H_{\mathrm{T}})_{\eta=+1}-(\delta H_{\mathrm{T}})_{\eta=-1}\right]/2$.

The above discussion lays the foundation of the extracting operations of i-DMI coefficient from precessional-flow cannyons 
under longitudinal in-plane fields, for example in Pt/Co/AlOx heterostructures 
by Thiaville and Pizzini \textit{et al.} in 2016\cite{Pizzini_EPL_2016,Thiaville_PRB_2016}.
However since they did not provide wall velocities under transverse in-plane fields, 
the existence of b-DMI can not be determined.

\begin{table} [htbp]
	\caption{Summary of $v\sim H_{\perp}$ canyons for field-driven DW dynamics in precessional flows.
		Definitions of $(\mathrm{p,q})$ and $H_{\mathrm{i(b)}}$ are the same as those in Table I.
		First and second row: Definition of $\tilde{f}_{\mathrm{p,q}}$ and $\tilde{g}_{\mathrm{p,q}}$.
		Third row: Unified $v\sim H_{\perp}$ expression.
		Fourth row: Unified maximum wall velocity at $H_{\perp}=\delta H_{\mathrm{p,q}}\pm\Delta H_{\mathrm{p,q}}$.
		Fifth row: Minimum wall velocity at canyon centers.
		Sixth row: Canyon center shifts (shaded for emphasization). 
		Last row: Half widths of canyons.}
	\renewcommand\arraystretch{1.8}
	\begin{tabular}  {p{40 pt}<{\centering} | p{48 pt}<{\centering} | p{45 pt}<{\centering} | p{45 pt}<{\centering} | p{48 pt}<{\centering}}
		\hline
		\hline
		$(\mathrm{p,q})$:  & $(\mathrm{i,L})$ & $(\mathrm{i,T})$ & $(\mathrm{b,L})$ & $(\mathrm{b,T})$   \\
		\hline
		$\frac{1}{\gamma}\cdot \tilde{f}_{\mathrm{p,q}}$:  & $H_z$ & $H_z +\frac{\eta\alpha\pi}{2} H_{\mathrm{i}}$  & $H_z-\frac{\eta\alpha\pi}{2}H_{\mathrm{b}}$ & $H_z$   \\
		\hline
		$\frac{2}{\alpha\pi\gamma}\cdot \tilde{g}_{\mathrm{p,q}}$:  & $H_{\mathrm{L}}+\frac{2}{\pi} H_{\mathrm{K}}-\eta H_{\mathrm{i}}$ & $H_{\mathrm{T}}-\frac{2}{\pi} H_{\mathrm{K}}$  & $H_{\mathrm{L}}+\frac{2}{\pi} H_{\mathrm{K}}$  & $H_{\mathrm{T}}-\frac{2}{\pi} H_{\mathrm{K}}-\eta H_{\mathrm{b}}$ \\
		\hline
		$v_{\mathrm{p,q}}$: &  \multicolumn{4}{c}{$\eta\frac{\Delta}{\alpha}\gamma H_z+\frac{\eta\Delta\cdot \tilde{f}_{\mathrm{p,q}}}{\alpha(1+\alpha^2)}\cdot\frac{|\tilde{g}_{\mathrm{p,q}}/\tilde{f}_{\mathrm{p,q}}|}{\ln\left(1-|\tilde{g}_{\mathrm{p,q}}/\tilde{f}_{\mathrm{p,q}}|\right)}$} \\
		\hline
		$v^{\mathrm{M}}$: & \multicolumn{4}{c}{$\eta\frac{\Delta}{\alpha}\gamma H_z$}  \\
		\hline
		$\frac{1+\alpha^2}{\eta\alpha\Delta\gamma}\cdot v_{\mathrm{p,q}}^{\mathrm{m}}$: & $H_z$ &  $H_z-\frac{\eta\pi}{2\alpha}H_{\mathrm{i}}$ & $H_z+\frac{\eta\pi}{2\alpha}H_{\mathrm{b}}$ & $H_z$  \\
		\hline
		\rowcolor{mygray}[1.82pt][1.82pt]  
		$\delta H_{\mathrm{p,q}}$: & $\eta H_{\mathrm{i}}-\frac{2}{\pi} H_{\mathrm{K}}$ & $\frac{2}{\pi} H_{\mathrm{K}}$  & $-\frac{2}{\pi} H_{\mathrm{K}}$ & $\eta H_{\mathrm{b}}+\frac{2}{\pi} H_{\mathrm{K}}$ \\
		\hline
		$\Delta H_{\mathrm{p,q}}$: & \multicolumn{4}{c}{$\frac{2}{\alpha\pi\gamma}|\tilde{f}_{\mathrm{p,q}}|$}  \\
		\hline
		\hline
	\end{tabular}
\end{table}

\section{V. Discussion}
Before the end of this paper, several points need to be clarified.
First, in the ``current-driven" scheme, the dome center shifts are independent on the real current density $j_{\mathrm{F}}$
flowing through the FM layers of magnetic heterostructures which is generally hard to directly measure.
This provide the universality of this scheme in determining the nature and strength of DMIs,
since it does not mix the intrinsic properties and external stimuli together.

Second, the strong in-plane current density and/or out-of-plane magnetic fields overcomes the pinning process 
and makes the precessional-flow mode of chiral DWs in longitudinal direction hardly affected by the stochastic fields 
originated from impurities and disorders in magnetic heterostructures. 
Also the relatively large wall velocity makes the experimental observation easier thus improve the data accuracy.
These are the extra advantages of our schemes except for their intrinsic universality.

Third, our theory holds under the assumption that $|g_{\mathrm{i(b),L(T)}}/f_{\mathrm{i(b),L(T)}}|<1$,
or equivalently not too far away from the dome summits or canyon bottoms.
Therefore, it can not explain the further evolution of wall velocity when in-plane fields go further beyond the half width
$\Delta H_{\mathrm{i(b),L(T)}}$.
Fortunately, the probing procedures of both DMIs [(C1)-(C5) or (F1)-(F5)] only depend on the position of 
dome summits or canyon bottoms, which makes our scheme universal.
Also, our theory holds for large enough in-plane currents or out-of-plane fields since 
now DWs precess almost evenly thus our linearization operation does not lose too much details of the entire circle.

At last, in our theory ``$q-\varphi-\Delta$" model\cite{He_EPJB_2013,jlu_PRB_2019} is adopted.
For ideal narrow-strip shaped heterostructures when considering the DMI-induced wall tiling $\chi$\cite{Boulle_PRL_2013} and canting $\theta_{\infty}$\cite{jlu_PRB_2016} in domains from in-plane fields, a more complicated wall ansatz
\begin{equation}\label{q_phi_chi_Delta_ansatz}
\tan\frac{\vartheta}{2}=\frac{e^R+\tan(\theta_{\infty}/2)}{1+e^R \tan(\theta_{\infty}/2)},\quad \phi=\varphi(t)
\end{equation}
can be proposed with $R\equiv \eta[(x-q)\cos\chi+y\sin\chi]/\Delta$.
By integrating the resulting dynamical equations over strip surface in $xy-$plane, 
alternative Lagrangian-based collective coordinate models, such as the ``$q-\varphi-\chi$"\cite{Boulle_PRL_2013} or ``$q-\varphi-\chi-\Delta$"\cite{Nasseri_JMMM_2017,Nasseri_JMMM_2018} models, emerge.
However they are too complicated to provide clear criteria in constructing operable procedures and explaining experimental data.
Generally in analyzing the position and shape 
of $v\sim H_{\perp}$ curves, the ``$q-\varphi-\Delta$" model is enough.
In addition, for real wider heterostructures with disorder, the walls take complex meander shape with its magnetization
vector rotating several times along the wall and thus show unconspicuous tilting $\chi$\cite{Marrows_PRB_2018,Pizzini_EPL_2016,Thiaville_PRB_2016}. 
This leads to negligible longitudinal (transverse) component of $\mathbf{H}_{\mathrm{b(i)}}$
which is proportional to $\nabla_y m_z$, 
hence explains the feasibility of procedures (C1)-(C5) and (F1)-(F5) for extracting both DMIs
in not-too-thin heterostructures.

\section{Acknowledgement}
J.L. acknowledges supports from Natural Science Foundation for Distinguished Young Scholars of
Hebei Province of China (A2019205310) and from National Natural Science Foundation of China (Grant No. 11374088).
M.L. is funded by the Project of Hebei Province Higher Educational Science and Technology Program (QN2019309).
X.R.W. is supported by the National Natural Science Foundation of China (Grants No. 11974296) and Hong
Kong RGC (Grants No. 16301518).


\end{document}